\documentclass[conference]{IEEEtran}
\IEEEoverridecommandlockouts

\usepackage{cite}
\usepackage{amsmath,amssymb,amsfonts}
\usepackage{algorithmic}
\usepackage{graphicx}
\usepackage{textcomp}
\usepackage{xcolor}
\usepackage{enumitem}
\usepackage{placeins}
\usepackage{float}
\usepackage{tabularx}
\usepackage{longtable}
\usepackage{booktabs}
\usepackage{multirow}
\usepackage{multicol}
\usepackage{caption}
\usepackage{subcaption}
\usepackage{array}
\usepackage{tabulary}
\usepackage{times}
\usepackage{titlesec}
\usepackage{hyperref}

\setlist[description]{leftmargin=\parindent,labelindent=\parindent}

\usepackage{url}

\usepackage{tikz}
\usetikzlibrary{positioning}
\usepackage{algorithm}
\usepackage[font=footnotesize]{caption}
\renewcommand{\figurename}{Figure~}
\usepackage[labelsep=period]{caption}

\def\BibTeX{{\rm B\kern-.05em{\sc i\kern-.025em b}\kern-.08em
    T\kern-.1667em\lower.7ex\hbox{E}\kern-.125emX}}

\begin{document}
\title{\textbf{\LARGE Constructing and Analyzing Different Density Graphs for Path Extrapolation in Wikipedia}} 
\author{\IEEEauthorblockN{Martha Sotiroudi, Anastasia-Sotiria Toufa, Constantine Kotropoulos}
\IEEEauthorblockA{Department of Informatics, Aristotle University of Thessaloniki
			\\ Thessaloniki, 54124, Greece \\
Email: \{marthass, toufaanast, costas\}@csd.auth.gr}
}
%

\maketitle

\begin{abstract}
Graph-based models have become pivotal in understanding and predicting navigational patterns within complex networks. Building on graph-based models, the paper advances path extrapolation methods to efficiently predict Wikipedia navigation paths.  The  Wikipedia Central Macedonia (WCM)  dataset is sourced from Wikipedia, with a spotlight on the Central Macedonia region, Greece, to initiate path generation. To build WCM, a crawling process is used that simulates human navigation through Wikipedia.  Experimentation shows that an extension of the graph neural network GRETEL, which resorts to dual hypergraph transformation, performs better on a dense graph of WCM than on a sparse graph of WCM. Moreover, combining hypergraph features with features extracted from graph edges has proven to enhance the model's effectiveness. A superior model's performance is reported on the WCM dense graph than on the larger \textsc{Wikispeedia} dataset, suggesting that size may not be as influential in predictive accuracy as the quality of connections and feature extraction. The paper fits the track 
Knowledge Discovery and Machine Learning of the 16th International Conference on Advances in Databases, Knowledge, and Data Applications.
\end{abstract}


\section{Introduction}
\label{sec:intro}

Graph structures offer an intuitive and powerful means to capture relationships and interactions within various kinds of data, paving the way for advanced analysis through the prism of Graph Neural Networks (GNNs) \cite{wu2020comprehensive,scarselli2008graph,zhang2019heterogeneous,zhou2020graph,kipf2016semi}.
From node classification \cite{xiao2022graph,li2019learning,rong2019dropedge} to link prediction \cite{kumar2020link} \cite{cai2021line}, GNNs have proven indispensable across a spectrum of applications. Among these, the task of link prediction focuses on path inference, namely to predict an agent's trajectory over a graph.

The efficacy of such models is inherently tied to the quality and structure of the underlying graph.
In this context, our work pivots on the creation of the Wikipedia Central Macedonia (WCM)  dataset, a  new dataset comprising paths extracted from the huge graph of Wikipedia, with a specific emphasis on articles related to Central Macedonia, Greece. The dataset tries to simulate human navigation paths as in \textsc{Wikispeedia}~\cite{wikispedia} game, where users are asked to navigate from a given source to a given target article by only clicking Wikipedia links. Our objective is to leverage this dataset to address the problem of path inference.

WCM  dataset is specifically designed to navigate through the complexities of Wikipedia’s topology. It takes ``Central Macedonia'' as the starting article, from which it explores the external links through a series of random walks. Each step is contingent on a set of well-defined validity criteria. This ensures that each selected link is pertinent and non-redundant, providing a true reflection of the path an agent might traverse within the bounds of this thematic cluster.
The dataset constructed for this study is made publicly available~\cite{WCM}. 
It comprises two separate files within the \texttt{Wikipedia\_Dataset} directory, representing the \texttt{Dense Graph} and the \texttt{Sparse Graph} structures, each containing details of the paths, unique articles, path identifiers, categories, edges, hyperedges, observations, and path lengths. The code to create the WCM dataset can be found at~\cite{github_repository}.

The interest in the path inference problem has led to the development of advanced models like GRETEL\cite{cordonnier2019extrapolating}, which has demonstrated promise in leveraging path extrapolation on graphs. GRETEL works as a generative model trying to capture the directionality of the path.
It has been applied to both navigation data and paths constructed on the Wikipedia graph.
This paper applies a graph transformation method based on the Dual Hypergraph Transformation (DHT)~\cite{jo2021edge}. This method, as demonstrated in \cite{toufa2023dual} \cite{toufa2023dualgretel+}, extends the traditional graph framework enabling connections among multiple nodes (i.e., vertices) within a hypergraph. Hypergraphs are suitable for this purpose because their edges can connect any number of nodes, not just two, as in a conventional graph. The new representation is able to capture more complex interactions between the data, and new more representative features can be extracted~\cite{Kotropoulos:15}.

Here, in pursuit of advancing path extrapolation methods, WCM dense and sparse graphs are employed to assess both the original GRETEL and the Dual GRETEL variant in environments of varying complexity, providing a thorough insight into its adaptability and accuracy in different graph densities. To capture a comprehensive range of interactions within the data, a feature extraction process is implemented as proposed in \cite{cordonnier2019extrapolating}\cite{toufa2023dual}\cite{toufa2023dualgretel+}. \cite{toufa2023dual} introduces an enhanced model, DualGRETEL+, that applies dual hypergraph transformation and a second-order optimizer to GPS navigation data, showing improved path inference capabilities. \cite{toufa2023dualgretel+} assesses path extrapolation using GRETEL on Wikipedia data, with a focus on extracting informative features through the DHT.

The paper is structured as follows: Section \ref{sec:methodology} provides a detailed description of the dataset creation and its characteristics, along with an overview of the features employed and the GRETEL model. A detailed exposition of the experiments and results is found in Section \ref{sec:experimetns}. The paper concludes in Section \ref{sec:conclusion}, underscoring the profound impact of graph density on the path extrapolation with graph neural networks.

\section{Methodology}
\label{sec:methodology}

This section focuses on the methodical approach to creating and analyzing the WCM, outlining the comprehensive process of collecting, categorizing, and extracting features from Wikipedia data to construct various graph types for path extrapolation.

\subsection{Dataset Creation}
The dataset is created through a crawling process designed to traverse the vast interconnected landscape of Wikipedia, with Wikipedia Central Macedonia article~\cite{WikipediaCentralMacedonia} serving as the focal starting point.
During data collection, we remained cognizant of the load implications on Wikipedia's servers. We inserted a pause of one second between two requests, safeguarding against potential server overload while accessing Wikipedia's data. This was a measure of digital courtesy and sustainability.

The path generation process begins with the Central Macedonia Wikipedia article. From this starting point, the crawler extracts all the external links associated with the current article. A subsequent article is then randomly selected from the set of external links, adhering to certain validity checks, ensuring the relevance of the link and its absence from the current path. To maintain the integrity of the dataset and concentrate solely on core articles, stringent validation criteria are instated.
The process of path creation continues until the generated path either attains a predetermined length ranging from 4 to 7 articles or encounters an article devoid of valid external links. The algorithm employs a well-defined criterion to ensure the relevance and validity of each article within the path. The function \texttt{is\_valid\_title} is utilized to exclude titles containing terms like \texttt{Talk}, \texttt{User}, \texttt{File}, `ISO', percentages, hashes, or colons, and those consisting solely of digits. This careful filtering is instrumental in maintaining a dataset focused on content-rich articles, avoiding disambiguation pages, meta-articles, or other forms of non-standard content that could detract from the dataset's integrity.

To ensure the intelligibility of the dataset, each Wikipedia article is associated with a distinct identifier. Leveraging tensor manipulation, the identifiers for the linked articles are distilled and organized within distinct tensor frameworks. These tensors serve as the foundation for the node indices within the constructed graph. To further aid our analysis, each trajectory's length is documented, and each article in the trajectory is associated with its unique identifier. This process is reiterated until a grand total of 3000 paths emerges.
The graph $G$ is created comprising $m$ nodes and $n$ edges, where nodes represent articles and edges denote links between articles. The extracted paths are referred to as \textit{trajectories}. We have documented these trajectories, noting their lengths and the articles they connect. The graph is represented using the Graph Markup Language.

Two distinct graph types have been created, each following a unique path selection process:
\begin{description}
    \item[Dense Graph]\hspace*{-1ex}: This is formed by a modified path selection protocol within the crawler. Here, the crawler opts for a random choice from the first five external links of an article. We choose the first five external links for path selection to intentionally narrow down the possible trajectories, aiming for a denser graph structure that facilitates a more focused analysis of interconnected topics. This results in a connected network among a smaller subset of 912 nodes, 1311 edges, and 3000 paths. The same process of path generation, involving the extraction of links, applying validity checks, and documenting each trajectory with unique identifiers, is followed as in the general dataset creation.

 \item[Sparse Graph]\hspace*{-1ex}: This graph follows the initial broader selection process, incorporating a more extensive set of 7307 nodes, 10612 edges, and 3000 paths. The selection is made from all the external links.
 
\end{description}

\subsection{Article Categorization}\label{AA}
Categorization provides a structured framework to analyze the dataset.  Organizing articles into distinct categories enables researchers to identify content trends and patterns within the generated paths. This categorization not only enriches the dataset but also amplifies its potential utility for diverse research, analytical, and educational purposes.

Our categorization strategy focuses on dynamic online querying using DBpedia \cite{auer2007dbpedia}. In order to determine the category of a given Wikipedia article, we rely on the SPARQL endpoint of DBpedia. Each article is queried to retrieve its semantic type from DBpedia's ontology. Whenever an explicit type is not obtained or if there are errors during the querying process, the articles are classified under \texttt{subject.General}.

\subsection{Feature Extraction}

In addition to graph generation, a feature extraction process is conducted to leverage semantic information from the content of the articles and to capture complex interactions in the graph structure.
According to \cite{cordonnier2019extrapolating}, the feature vector for the nodes corresponds to its \texttt{in/out degree}, and its length is 2. For edges, the feature vector contains the Text Frequency - Inverse Document Frequency (\texttt{TF-IDF score}), capturing the semantic similarity between source and destination articles of a hyperlink \cite{ramos2003using}, and the number of times the link was clicked in the training dataset of paths (\texttt{nof}).

\subsubsection{Dual Hypergraph Transformation}

The framework commences with the configuration of a conventional graph, designated as $G$ having $n$ nodes and $m$ edges. Node features are represented by a feature matrix $ \pmb{F} \in \mathbb{R}^{n \times d} $, and edge features by a feature matrix $ \pmb{E} \in \mathbb{R}^{m \times d'}$. Here, $ d $ and  $d'$ are the size of node and edge feature vectors, respectively.
Considering an undirected graph, the incidence matrix is defined as $ \pmb{M} \in \{0, 1\}^{n \times m}$. In the case of a directed graph, the incidence matrix is defined as $\pmb{M} \in \{-1,0,1\}^{n \times m}$. In any case, the incidence matrix represents the relationships between nodes and edges in a graph, indicating which nodes are connected by specific edges.

The conventional graph and the corresponding dual hypergraph are represented as $G = (\pmb{F}, \pmb{M}, \pmb{E})$ and $G^\ast = (\pmb{F}^\ast, \pmb{M}^\ast, \pmb{E}^\ast)$ respectively. $\pmb{F}^\ast$ represents the node features of hypergraph while $\pmb{E}^\ast$ represents the hyperedge features. The DHT algorithm interchanges the roles of nodes and edges of the original graph \cite{jo2021edge}. That is, the edges of the original graph are reinterpreted as nodes in the dual hypergraph, while the original nodes become hyperedges in the dual hypergraph. Accordingly,  $\pmb{F}^\ast = \pmb{E} \in \mathbb{R}^{m \times d'}$ and $\pmb{E}^\ast = \pmb{F} \in \mathbb{R}^{n \times d} $. The incidence matrix of the dual hypergraph is the transpose of the incidence matrix of the original graph, i.e., $\pmb{M}^\ast = \pmb{M}^\top$.
The transformation is mathematically defined as:
\begin{equation}
G = (\pmb{F}, \pmb{M}, \pmb{E}) \rightarrow G^\ast = (\pmb{E}, \pmb{M}^\top, \pmb{F})
\end{equation}
Notably, the DHT is a reversible transformation, ensuring that applying it to $G^\ast$ recaptures the initial graph $G$, thereby preserving the structural and feature integrity of the transformation.

\subsubsection{Features extracted from the dual hypergraph}

 Following the methodology proposed in \cite{toufa2023dual}, the original graph is transformed into its corresponding dual graph by applying the DHT algorithm in order to capture more complex interactions among edges. 
 Two new features are extracted, namely the \texttt{similarity-hyperedge} and the \texttt{DHnode-in-out-degree}.  
 The first feature assumes an undirected graph, while the second one assumes a directed graph. The implementation of dual hypergraph feature extraction, which significantly enhances the predictive accuracy of our models, can be found in~\cite{Wikispedia_GRETEL}.
   
For the \texttt{similarity-hyperedge} feature, the first step is to construct the incidence matrix $\pmb{M} \in \{0,1\}^{n\times m}$. Row vector $\pmb{q}_l \in \{0,1\}^m$ of $\pmb{M}$, corresponds to node $l$.  The cosine similarity between the incidence row vectors $\pmb{q}_v$ and $\pmb{q}_u$ is computed, where $v$ is the source node and $u$ is the target node of an arbitrary edge $e$.
The corresponding vector in the $\pmb{M}^\ast$ matrix is a column vector $\pmb{q}_l^{\ast} \in \{0,1\}^m \equiv \pmb{q}_l^\top$. The position of each 1 in this column vector indicates which nodes of the dual hypergraph are connected with the hyperedge $l^\ast$.

For the \texttt{DHnode-in-out-degree}, a directed graph $G$ is assumed.  The corresponding incidence matrix is defined as $\pmb{M} \in \{-1,0,1\}^{n\times m}$. To extract features associated with the input and output degrees of the dual hypergraph nodes, determining the direction of hypergraph edges becomes essential.
This involves an examination of the column vector of $\pmb{M}^\ast$ $\pmb{q}_l^{\ast} \equiv \pmb{q}_l^\top$. The position of each 1 in this column vector indicates which nodes of the dual hypergraph are connected with the hyperedge $l^\ast$. For every combination $(v^\ast_i, v^\ast_j)$, we verify the existence of a path $e_i \rightarrow e_j$ in the original graph that passes through the scrutinized node $l$. The new feature is the in-degree and out-degree of dual hypergraph nodes which are normalized by the maximum observed degree $D_{\max }$ in the hypergraph to facilitate comparison across different nodes:
\begin{equation}
\resizebox{0.9\hsize}{!}{$
\text {\scriptsize Normalized In/Out-Degree }\left(v_i^*\right)=\frac{\text {In/Out-Degree }\left(v_i^*\right)}{D_{\max }}$}
\end{equation}
The aggregation of \texttt{similarity-hyperedge} and \texttt{DHnode-in-out-degree} results in \texttt{Similarity-Hyperedge-DHnode-In-Out-Degree} feature. These enhanced features are particularly critical in the sparse graph context, where the reduced number of connections demands a more nuanced approach to capturing node relationships. In the dense graph, with its inherently richer connectivity, these features play a pivotal role in distilling the essence of the network's complexity into a format conducive to advanced path prediction algorithms.

The feature extraction procedure is performed on the sparse graph with 7,307 nodes and 10,611 edges and the dense graph with 912 nodes and 1,311 edges.

\subsection{Path Extrapolation Employing GRETEL}

The paper addresses path extrapolation focusing on predictive path analysis via the GRETEL model \cite{cordonnier2019extrapolating}. The graph $G$ consists of nodes and edges, represented as $G=(\mathcal{V}, \mathcal{E})$, with $n=|\mathcal{V}|$ denoting the node count and $m=|\mathcal{E}|$ the edge count, respectively. An agent progresses through the graph, stepping from node $v_i$ to $v_j$ contingent on the presence of a directed edge $e_{i \rightarrow j} \in \mathcal{E}$.

The agent's position at time $t$ is a sequential set of traversed nodes, symbolized as a given prefix $ p=\left(v_1, v_2, \ldots, v_t\right)$.  Let the path suffix  $s=\left(v_{t+1}, \ldots, v_{t+h}\right)$ be a collection of potential future for  prediction horizon $h$. Within this setting, GRETEL is leveraged to estimate the conditional likelihood
$\operatorname{Pr}(s\mid h, p, G)$ of path suffix $s$ given the prefix $p$, the horizon $h$, and the graph $G$.  The agent's position at each step $t$ is encoded by a sparse vector $\mathbf{x}_t \in \mathbb{R}_{\geq 0}^n$  normalized to a unit sum, with its $i$-th element reflecting the likelihood of the agent being at node $v_i$.

GRETEL constructs a generative model that considers the directionality of edges via a latent graph with edge weights informed by a Multi-Layer Perception (MLP) that respects the graph's inherent directionality. The model's essence lies in its ability to forecast paths by learning from the traversed sequences, leveraging node features and the collective path history. More specifically, the non-normalized weights of each edge are computed by
\begin{equation}
    z_{i\to j} = \mathrm{MLP} (\pmb{c}_i, \pmb{c}_j, \pmb{f}_i, \pmb{f}_j, \pmb{f}_{i\to j}),
    \label{Eq:MLP}
\end{equation}
where $\pmb{c}_i$ and $\pmb{c}_j$ are the pseudo-coordinates of the sender and the receiver node, respectively, while $\pmb{f}_i$ and $\pmb{f}_j$ denote the features of the sender and the receiver node, respectively. In (\ref{Eq:MLP}), $\pmb{f}_{i\to j}$ is the feature vector of the edge that connects the sender and the receiver node. The computed MLP outputs are normalized with the softmax function.
The pseudo-coordinates $\pmb{c}_i$  are computed using a GNN of $K$ layers. They are the agent representations $\pmb{x}_{\tau}$ for $\tau \in \mathcal{I}$, where $\mathcal{I}$ denotes a trajectory. The non-zero elements of $\pmb{x}_{\tau}$ refer to the distance between the agent and the $K$ closest graph nodes normalized to measure one.
Let $\vec{e}_t$ and $\vec{e}_t$ define the edges that go from $v_t \to v_{t+1}$ and $v_t \to v_{t-1}$, respectively. Let also $\pmb{x}_t$ be the last position of the agent. GRETEL~\cite{cordonnier2019extrapolating} can be trained through the \textit{target likelihood}. That is, given a target distribution $\pmb{x}_{t+h}$, the model tries to estimate the destination distribution $\hat{\pmb{x}}_{t+h} \in \mathbb{R}^{n \times 1}$ over a horizon $h$  by the non-backtracking walk~\cite{kempton2016non}
\begin{equation}
    \hat{\pmb{x}}_{t+h} = \pmb{B}_{\phi}^{+} \pmb{P}_{\phi}^h \pmb{B}_{\phi} \pmb{x}_t.
    \label{Eq:x(t+h)}
\end{equation}
Let $w_{\phi} (e_{k \to j})$ stand for the normalized MLP weights. In (\ref{Eq:x(t+h)}), $\pmb{P}_\phi \in \mathbb{R}^{m \times m}$ has elements
\begin{equation}
[\pmb{P}_\phi]_{e_{i \to j}, e_{k \to l}} =
\begin{cases}
0 & \text{if } j \neq k \text{ or } i=l \\
\frac{w_\phi ({e}_{k \to l})}{1 - w_\phi (e_{k \to i})} & \text{otherwise,}
\end{cases}
\end{equation}
$\pmb{B}_\phi$ is a $m \times n$ matrix with $[\pmb{B}_\phi]_{e_{i\to j}, k} = 0$ if $k \neq i$ and $w_{\phi} (e_{k \to j})$, otherwise, and  $\mathbf{B}_{\phi}^{+}$ stands for the pseudoinverse of $\mathbf{B}_{\phi}$.
Such an approach integrates node and edge feature vectors, the former delineating the in/out-degree and the latter embedding the textual and usage-based similarity metrics. These primal features are pivotal in the model's capacity to estimate the suffix likelihood, aiding in approximating the path probability $\operatorname{Pr}(s \mid h, \varphi, G)$. In the paper, we will aggregate the original edge features $\pmb{f}_{i\to j}$ with the features extracted from the dual hypergraph.



\section{Experiments and Results}
\label{sec:experimetns}

To quantify the structure of each graph, we calculate the density, which provides a measure of how complete the graph is. The density is defined as the ratio of the number of edges $m$ to the number of possible edges, with the formula given for a directed graph without loops as
\begin{equation}
D = \frac{m}{n(n-1)},
\end{equation}
where $n$ is the number of nodes.

\begin{table}[!ht]
\caption{\sc Dataset Characteristics}
\centering
\small
\resizebox{0.35\textwidth}{!}{%
\begin{tabular}{|l|c|c|c|}
\hline
\textbf{Datasets} & \textbf{Nodes} & \textbf{Edges} & \textbf{Density} \\
\hline
\textbf{Sparse Graph} &7307 &10612 &\(2 \times 10^{-4}\)\\
\hline
\textbf{Dense Graph} &912 &1311 &\(1.58 \times 10^{-3}\) \\
\hline
\textbf{Wikispeedia} &4604 &119882 &\(5.66 \times 10^{-3}\) \\
\hline
\end{tabular}}
\label{tab:dataset_characteristics}
\end{table}

Table \ref{tab:dataset_characteristics} summarizes the characteristics of the graphs used in the experiments, providing a clear comparison of the number of nodes, edges, and density across the Sparse Graph, the  Dense Graph, and \textsc{Wikispeedia}.
Based on the characteristics outlined in Table~\ref{tab:dataset_characteristics}, the sparse graph demonstrates a lower density ratio due to its larger node count. In contrast, the dense graph, with fewer nodes, exhibits a higher density ratio. Notably, the \textsc{Wikispeedia} dataset possesses the greatest density ratio of the three.

The following metrics are used to assess the feature vectors. \emph{Target probability}  measures the average chance that the model will choose a node with non-zero likelihood. \emph{Choice accuracy} measures how accurate the decisions of an algorithm are at each crossroad of the ground-truth path, connecting nodes $v_t$ and $v_{t+h}$. It is computed on nodes whose degree is at least 3. \emph{precision top1} measures how often the correct next step appears in the model's first prediction only, while \emph{precision top5} evaluates how often the correct next step appears within the model's first five predictions.

In all experiments, the node feature vector includes the in/out degree for the nodes, retaining a constant size of two, underscoring consistent complexity in nodal characteristics despite the variation in graph densities.

An empirical assessment of model performance using the features derived from the original graph and those of the corresponding dual hypergraph is conducted.  In the case of \texttt{original edges}, the \texttt{TF-IDF score} and \texttt{nof} features are used, yielding a feature vector of size 2. By aggregating the features  \texttt{similarity-hyperedge} of length 1,  \texttt{ DHnode-in-out-degree} of length 2, and their combination \texttt{similarity-hyperedge-DHnode-in-out-degree} of length 3, the associated edge feature vector has length 3, 4, and 5, respectively.

\begin{table*}[!ht]
\caption{\sc Performance Metrics (\%) on the Sparse Graph}
\centering
\resizebox{0.9\textwidth}{!}{%
\begin{tabular}{|c|c|p{8em}|p{8em}|p{10em}|}
\hline
&
\multicolumn{1}{|c|}{GRETEL} &
\multicolumn{3}{c|}{Dual GRETEL} \\
\hline
\textbf{Metrics} & \textbf{Original Edges} & \textbf{Similarity-Hyperedge} &  \textbf{DHnode-In-Out-Degree} & \textbf{Similarity-Hyperedge-DHnode-In-Out-Degree}\\
\hline
\textbf{target probability} & $68.76 \pm 0.0044$ & $68.76 \pm 0.0019$ & $68.99 \pm 0.0064$ & $\mathbf{69.71 \pm 0.0038}$\\
\hline
\textbf{choice accuracy} & $\mathbf{51.18 \pm 0.0011}$ & $38.69 \pm 0.0042$ & $39.60 \pm 0.0082$  & $39.24 \pm 0.0090$\\
\hline
\textbf{precision top1} & $66.65 \pm 0.0050$ & $66.71 \pm 0.0012$ & $66.65 \pm 0.0025$ & $\mathbf{67.14 \pm 0.0045}$\\
\hline
\textbf{precision top5} & $80.62 \pm 0.0019$ & $80.62 \pm 0.0019$  & $80.68 \pm 0.0023$ & $\mathbf{80.98 \pm 0.0036}$\\
\hline
\end{tabular}}
\label{tab:sparse_metrics}
\end{table*}

\begin{table*}[!ht]
\caption{\sc Performance Metrics (\%) on the Dense Graph}
\centering
\resizebox{0.9\textwidth}{!}{%
\begin{tabular}{|c|c|p{8em}|p{8em}|p{10em}|}
\hline
&
\multicolumn{1}{|c|}{GRETEL} &
\multicolumn{3}{c|}{Dual GRETEL} \\
\hline
\textbf{Metrics} & \textbf{Original Edges} & \textbf{Similarity-Hyperedge}& \textbf{DHnode-In-Out-Degree} & \textbf{Similarity-Hyperedge-DHnode-In-Out-Degree} \\
\hline
\textbf{target probability} & $0.0030 \pm 0.0021$ & $\mathbf{19.1007 \pm 0.0004}$ &  $18.8741 \pm 0.0033$ & $19.0980 \pm 0.0026$\\
\hline
\textbf{choice accuracy} & $\mathbf{48.0602 \pm 0.0135}$ & $27.8261 \pm 0.0084$  & $29.8662 \pm 0.0096$& $29.5318 \pm 0.0086$\\
\hline
\textbf{precision top1} & $0.001 \pm 0.0023$ & $19.8995 \pm 0.0074$  & $18.0904 \pm 0.0075$& $\mathbf{20.5025 \pm 0.0067}$\\
\hline
\textbf{precision top5} & $0.2513 \pm 0.0012$ & $83.2161 \pm 0.0088$  & $82.8141 \pm 0.0258$& $\mathbf{83.8694 \pm 0.0112}$\\
\hline
\end{tabular}}
\label{tab:dense_metrics}
\end{table*}

\begin{table*}[!ht]
\caption{\sc Performance Metrics (\%) on the \textsc{Wikispeedia} Dataset}
\centering
\resizebox{0.9\textwidth}{!}{%
\begin{tabular}{|c|c|p{8em}|p{8em}|p{10em}|}
\hline
&
\multicolumn{1}{|c|}{GRETEL} &
\multicolumn{3}{c|}{Dual GRETEL} \\
\hline
\textbf{Metrics} & \textbf{Original Edges} & \textbf{Similarity-Hyperedge} & \textbf{DHnode-In-Out-Degree}& \textbf{Similarity-Hyperedge-DHnode-In-Out-Degree} \\
\hline
\textbf{target probability} & $6.42 \pm 0.1$ & $\mathbf{6.74 \pm 0.1}$  & $6.44 \pm 0.2$ & $6.2 \pm 0.1$\\
\hline
\textbf{choice accuracy} & $22.16 \pm 0.4$ & $\mathbf{23.2 \pm 0.1}$  & $22.88 \pm 0.1$ & $21.86 \pm 0.4$\\
\hline
\textbf{precision top1} & $11.6 \pm 0.2$ & $\mathbf{12.7 \pm 0.1}$  & $12.14 \pm 0.1$ & $11.66 \pm 0.3$\\
\hline
\textbf{precision top5} & $30.1 \pm 0.1$ & $\mathbf{30.14 \pm 0.1}$  & $30.02 \pm 0.05$ & $30 \pm 0.09$\\
\hline
\end{tabular}}
\label{tab:wikispeedia_metrics}
\end{table*}

Table~\ref{tab:sparse_metrics} summarizes the performance of the GRETEL model with original edge features and the features extracted from the dual hypergraph (Dual GRETEL) added on top of the original edge features on the Sparse Graph.

Table~\ref{tab:dense_metrics} repeats the model's performance assessment on the Dense Graph.
Table~\ref{tab:wikispeedia_metrics} details the model's performance on the \textsc{Wikispeedia} dataset. This dataset encapsulates the essence of human navigational strategies within Wikipedia, compiling 51318 completed paths from the \textsc{WIKI GAME} where participants navigate through article links towards a target article, with an aim for efficiency in both clicks and time. 

The modularity class algorithm in \texttt{Gephi} \cite{Gephi} is used to identify the clusters within the network. These clusters contain nodes that are more densely connected to each other than to nodes in different clusters. 
The resulting clusters are indicated by the color coding of the nodes. The size of each node is proportional to its degree, reflecting the number of connections it has within the network. This allows for the immediate visual identification of highly connected nodes. The visible labels on the nodes in the figures were chosen because they have higher degree values, which show their importance in the graph, and they represent the main topic of each cluster within the expansive Wikipedia network.

\renewcommand{\figurename}{Figure}
\begin{figure*}[!ht]
  \centerline{\includegraphics[width=14cm]{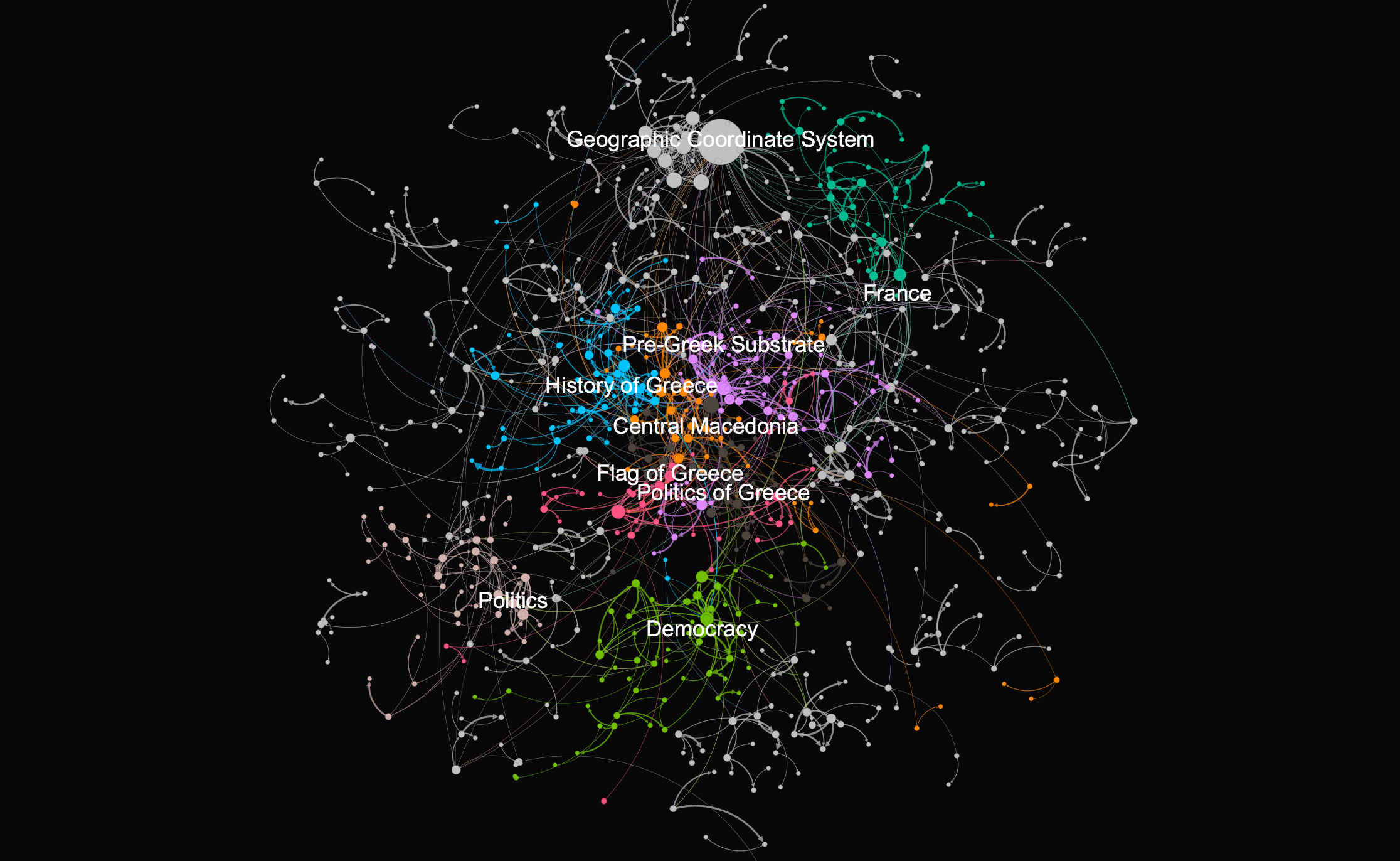}}
  \caption{Dense Wikipedia Graph.}
  \label{dense_graph}
\end{figure*}

Figure~\ref{dense_graph} represents the dense graph of Wikipedia. The selective navigation results in a dense network with several clusters, one of which is built around the Central Macedonia article, connecting closely related topics. Adjacent nodes like `History of Greece' and `Politics of Greece' form clusters that delve into the nation’s past and governance, and `Geographic Coordinate System' and `France' appear as nodes indicative of broader geographical discourse.

\begin{figure*}[!ht]
  \centerline{\includegraphics[width=14cm]{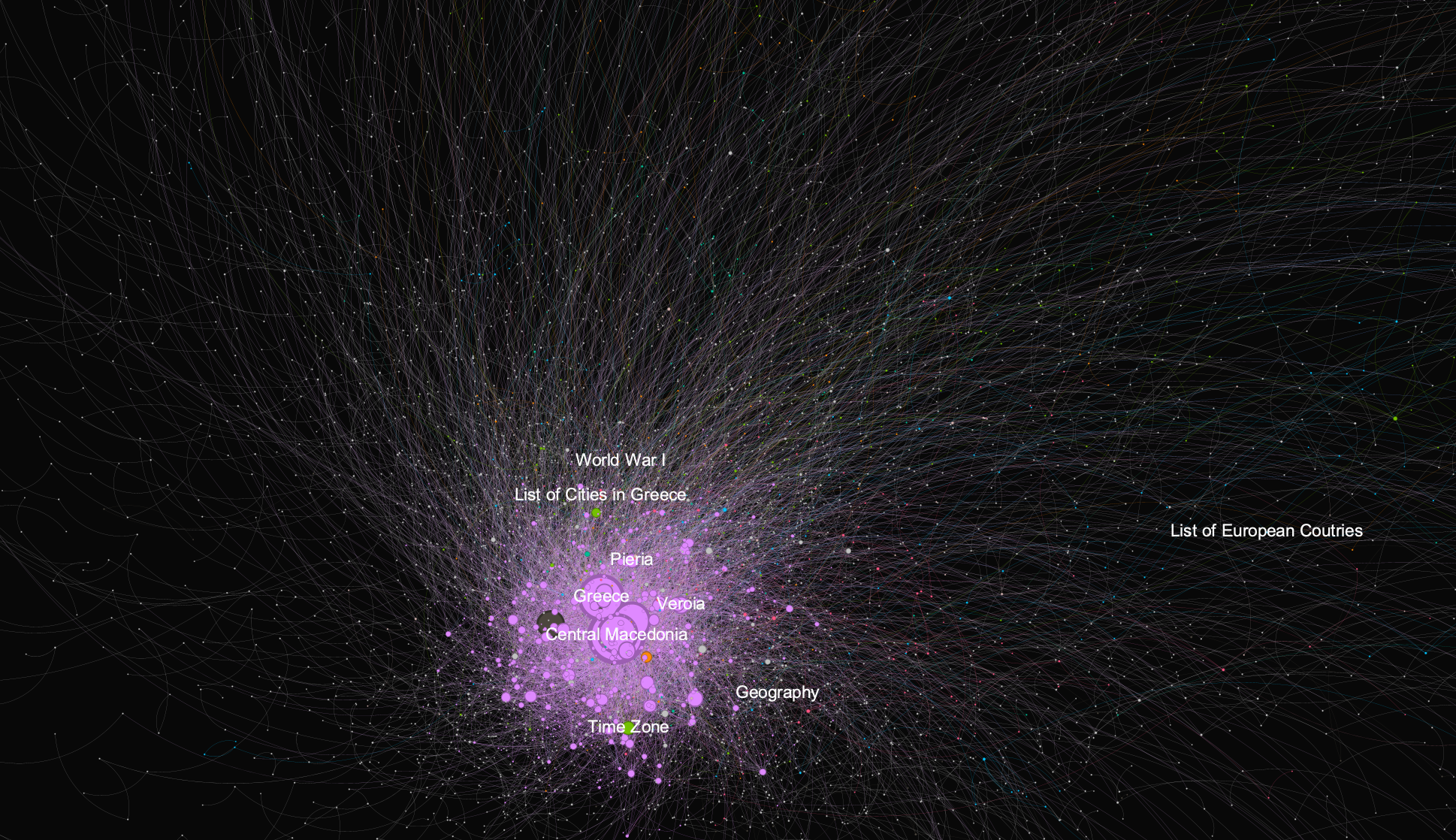}}
  \caption{Sparse Wikipedia Graph.}
  \label{sparse_graph}
\end{figure*}

\begin{figure*}[!ht]
  \centerline{\includegraphics[width=14cm]{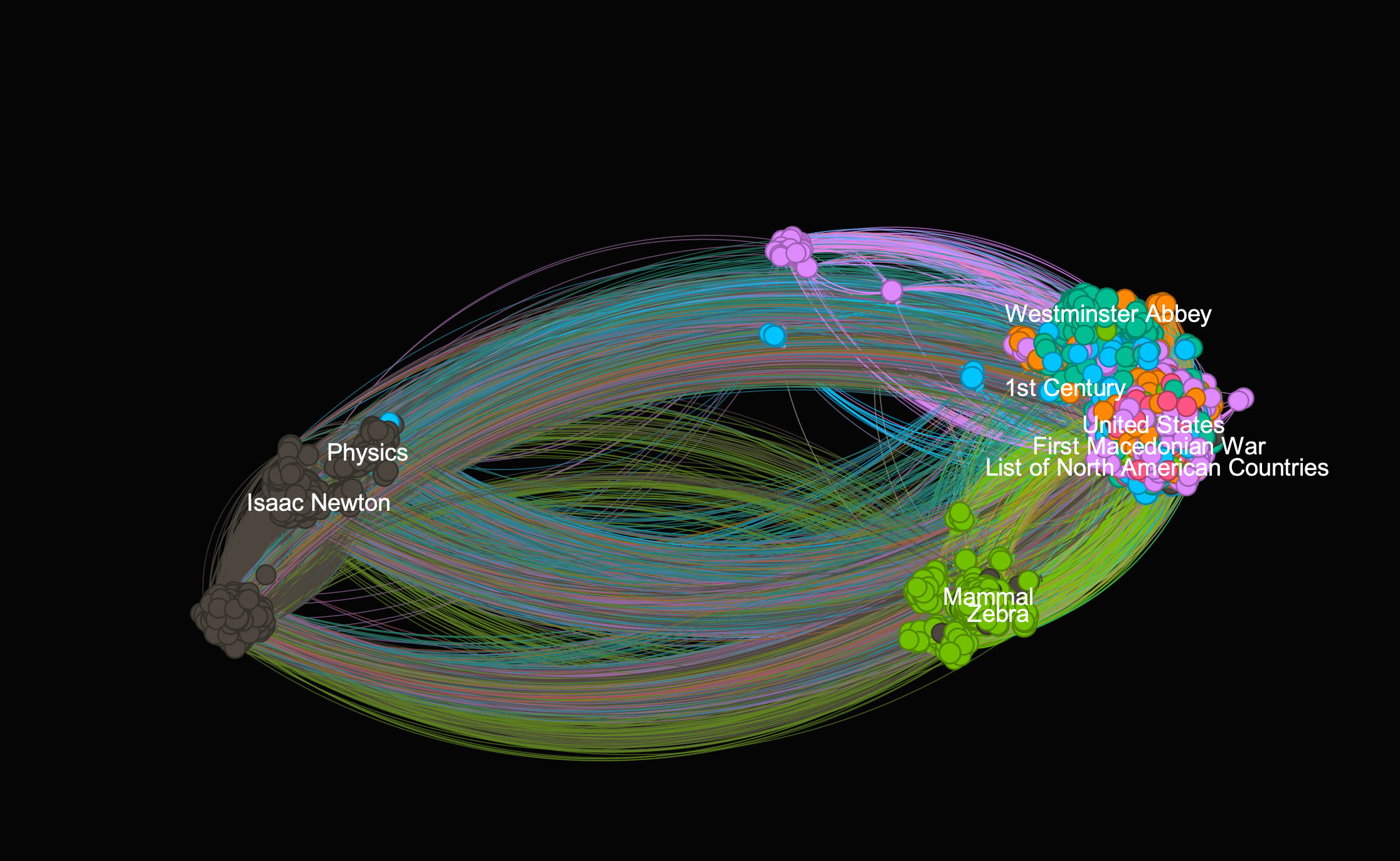}}
  \caption{Wikispeedia Graph.}
  \label{wikispeedia_graph}
\end{figure*}

The visualization of the sparse graph in Figure~\ref{sparse_graph} reveals a network that unfolds from the Central Macedonia article, forming a large, primary cluster due to the random link selection strategy, and extending outward into a sparse array of smaller clusters. These smaller clusters are thematic, with subjects such as European countries, Greek cities, and historical events.

The construction methods of the two graphs distinctly shape their representations. The dense graph demonstrates that the Central Macedonia article forms a cluster, with surrounding clusters closely related in theme, predominantly focusing on Greece. This clustering suggests that the used method tends to group related topics tightly together. On the other hand, the sparse graph shows a different pattern where the Central Macedonia article and closely linked articles stand out in number, while other articles appear less connected. This difference highlights how the choice of links in the construction process can significantly affect the network's structure.

Figure~\ref{wikispeedia_graph} represents the Wikispeedia graph, characterized by uniformly sized nodes, indicative of a network without a predominant starting article. Clusters within the graph are thematically organized, with `Isaac Newton' and `Physics' forming a cluster around scientific inquiry, while `Westminster Abbey' serves as a node for the cluster concerning England. `Mammal' and `Zebra' are central to a cluster on zoology. These labels serve as the focal points for their respective clusters, marking the diverse subjects navigated by users.

Table~\ref{tab2} showcases examples of how the extracted features are employed to predict specific paths, highlighting the model’s ability to deduce the most probable outcomes. Table~\ref{tab2} includes the conditional probabilities that reflect the model's ability to correctly anticipate the actual path taken. These examples are instrumental in illustrating the practical application of the model and the effectiveness of the features in guiding the model toward the most probable navigational route. The utilization of hypergraph features results in higher conditional probability compared to the use of \texttt{original edge} features. The examples clearly show that when hypergraph features are considered, the model tends to assign a greater likelihood to the true path, suggesting that these features capture more of the complexities inherent in human navigational behaviors on Wikipedia. The examples are drawn from the sparse graph of the WCM dataset.
 
\begin{table*}[ht]
\caption{\sc Examples of Path Prediction}
 \centering
 \resizebox{\textwidth}{!}{%
 \begin{tabular}{|l|l|l|l|l|l|l|}
\hline
 &  & \textbf{$\Pr (s \mid h, p, G)$} &  & \textbf{$\Pr (s \mid h, p, G)$} &  & \textbf{$\Pr (s \mid h, p, G)$} \\ \hline
\textbf{prefix} & \textbf{\begin{tabular}[c]{@{}l@{}}Naousa, Imathia, \\History of Macedonia, Craterus\end{tabular}} &  & \textbf{\begin{tabular}[c]{@{}l@{}} Volvi, Egnatia, \\Thessaloniki, Arethousa\end{tabular}} &  & \textbf{\begin{tabular}[c]{@{}l@{}}Thessaloniki, Greek National Road, \\Evzonoi, Axioupoli  \end{tabular}} &  \\ \hline
\textbf{true suffix} & \textbf{Antigenes, Nearchus, Tlepolemus} &  & \textbf{Nea Madytos, Vrasna} &  & \textbf{Greek Macedonia, Despotate of Epirus} &  \\ \hline
\textbf{original edges} & \begin{tabular}[c]{@{}l@{}} Antigenes, Nearchus, Satraps\\ \textbf{Antigenes, Nearchus, Tlepolemus}\end{tabular} & \begin{tabular}[c]{@{}l@{}}0.74\\ 0.26\end{tabular} & \begin{tabular}[c]{@{}l@{}}Stefanina, Thessaloniki\\ \textbf{Nea Madytos, Vrasna }\end{tabular} & \begin{tabular}[c]{@{}l@{}}0.75 \\ 0.25\end{tabular} &\begin{tabular}[c]{@{}l@{}}
Skra, Kilkis \\ \textbf{Greek Macedonia, Despotate of Epirus} \end{tabular} & \begin{tabular}[c]{@{}l@{}}0.38 \\ 0.01 \end{tabular}\\ \hline
\textbf{similarity-hyperedge} & \begin{tabular}[c]{@{}l@{}} Antigenes, Nearchus, Satraps\\ \textbf{Antigenes, Nearchus, Tlepolemus}\end{tabular} & \begin{tabular}[c]{@{}l@{}}0.64\\ 0.36\end{tabular} &
\begin{tabular}[c]{@{}l@{}}Stefanina, Thessaloniki\\ \textbf{Nea Madytos, Vrasna }\end{tabular}
& \begin{tabular}[c]{@{}l@{}}0.67\\ 0.33\end{tabular}&
\begin{tabular}[c]{@{}l@{}}
Skra, Kilkis \\ \textbf{Greek Macedonia, Despotate of Epirus} \end{tabular}
 & \begin{tabular}[c]{@{}l@{}} 0.26\\ $\mathbf{0.03}$ \end{tabular}
\\ \hline
\textbf{DHnode-in-out-degree} & \begin{tabular}[c]{@{}l@{}} Antigenes, Nearchus, Satraps\\ \textbf{Antigenes, Nearchus, Tlepolemus}\end{tabular} & \begin{tabular}[c]{@{}l@{}}0.69\\ 0.31\end{tabular} &
\begin{tabular}[c]{@{}l@{}}Stefanina, Thessaloniki\\ \textbf{Nea Madytos, Vrasna }\end{tabular}
& \begin{tabular}[c]{@{}l@{}}0.78\\ 0.22 \end{tabular}&
\begin{tabular}[c]{@{}l@{}}
Skra, Kilkis \\ \textbf{Greek Macedonia, Despotate of Epirus} \end{tabular}
& \begin{tabular}[c]{@{}l@{}} 0.29 \\ 0.01\end{tabular}\\ \hline
\textbf{\begin{tabular}[c]{@{}l@{}}similarity-hyperedge\\ - DHnode-in-out-degree\end{tabular}} & \begin{tabular}[c]{@{}l@{}} \textbf{Antigenes, Nearchus, Tlepolemus}\end{tabular} & \begin{tabular}[c]{@{}l@{}} $\mathbf{0.6}$ \end{tabular} &
\begin{tabular}[c]{@{}l@{}}Stefanina, Thessaloniki\\ \textbf{Nea Madytos, Vrasna }\end{tabular}
 & \begin{tabular}[c]{@{}l@{}}0.58\\ $\mathbf{0.42}$\end{tabular}
 &   \begin{tabular}[c]{@{}l@{}}
Skra, Kilkis  \end{tabular} & \begin{tabular}[c]{@{}l@{}} 0.46 \end{tabular} \\ \hline
\end{tabular}%
}
\label{tab2}
\end{table*}

\subsection{Performance Analysis of Model Across Dense and Sparse Graphs}

In the evaluation of the Dual GRETEL model, distinct performances are observed between the sparse and dense graphs. A higher predictive accuracy with respect to \emph{precision top5} metric is measured for the dense graph than the sparse one. This improved performance can be attributed to the vital role of the hyperedges, which enrich the model's contextual framework for more accurate extrapolation.

The sparse graph, despite its lower connectivity, shows commendable results, outperforming the dense graph in terms of target probability and choice accuracy. 
Dual GRETEL predicts the correct target with a probability of 69.71 $\pm$ 0.0038 \%. GRETEL accurately chooses the next step with a rate of 51.18 $\pm$ 0.0011 \%.   It's noteworthy that except for the precision top 5, Dual GRETEL maintains a better performance on the sparse graph than the dense one.

This comparison reveals that while the Dual GRETEL model benefits from the rich link structures in dense graphs for precision tasks, it retains substantial predictive strength in sparse settings. This insight may guide further optimization for the model, enhancing its adaptability across varying network densities.

 Also, this performance indicates that the model may benefit from the reduced complexity in sparse networks, potentially due to less noise and fewer connections, which can simplify the path prediction process. The comparison suggests that the model might generalize better in sparse environments, avoiding potential overfitting that can occur in dense networks with more intricate connections. Conversely, the specificity that dense networks provide can enhance the model's precision in certain contexts.

 \subsection{Model Benchmarking on WCM Dense Graph Versus Wikispeedia Graph}

For the WCM dense graph, Dual GRETEL demonstrated a significant improvement, achieving an impressive precision top5 score of 83.8694 $\pm$ 0.0112\%. On the WIKISPEEDIA graph, Dual GRETEL also showed enhanced performance with a precision top5 score of 30.14 $\pm$ 0.1\%. This indicates that hypergraph features greatly enhance the model's ability to accurately identify the most likely paths in a dense environment.

Furthermore, GRETEL demonstrated a high choice accuracy of 48.0602 $\pm$ 0.0135\% on the dense graph compared to 23.2 $\pm$ 0.1\% of Dual GRETEL on the \textsc{Wikispeedia} dataset. Our findings show that model performance on the dense graph improves across all metrics except \emph{choice accuracy} when we use hypergraph features.
That is, hypergraph features are particularly effective in densely connected graphs, enhancing the model's predictive accuracy across all metrics we tested. The results indicate the potential of hypergraph features to improve the performance of path prediction models like GRETEL, especially in complex network structures.

The completion rate of paths in the \textsc{Wikispeedia} dataset may introduce additional complexity, given that there is a mixture of successful and abandoned paths. In contrast, the smaller dataset might offer more uniformly successful paths, influencing the ease with which the model can learn and predict.

 The analysis of the model's performance, as shown in Tables~\ref{tab:sparse_metrics}-\ref{tab:wikispeedia_metrics}, reveals a trend where effectiveness inversely correlates with graph density. This suggests that as graphs become more interconnected, the model encounters greater challenges in path prediction accuracy. These observations emphasize the critical role that graph density plays in the deployment and refinement of path prediction algorithms. A possible explanation for the deterioration of accuracy as density increases could be the rise in potential paths that the model must discern. In denser graphs, the increased interconnectivity results in a greater number of plausible trajectories between nodes, potentially complicating the model's task of pinpointing the most likely path. Furthermore, a dense network may introduce more noise in the form of less relevant or weaker connections, which could mislead the prediction algorithm. These findings indicate that models like GRETEL or Dual GRETEL may require adjustments or enhancements, such as more sophisticated feature extraction or the incorporation of context-aware learning mechanisms, to better handle the complexity introduced by higher-density graphs.

\section{Conclusions}
\label{sec:conclusion}

A detailed analysis of GRETEL and its variant Dual GRETEL has been presented on dense and sparse graphs derived from the WCM dataset, aiming to improve path extrapolation models. Having developed the novel dataset centered on Central Macedonia, Greece, we have provided a resource that captures the complexity of human navigational patterns on Wikipedia.

Our investigation has shown that the density of a graph significantly influences the effectiveness of path prediction methods. Both models have performed better on sparse graphs in various aspects, yet they have achieved higher accuracy with respect to the top five predictions on the dense WCM graph. Furthermore, the incorporation of hypergraph features into the GRETEL model yielding the Dual GRETEL variant has significantly enhanced the accuracy of path predictions, underscoring the importance of feature extraction in graph-based predictive analytics. Comparisons of Dual GRETEL performance on the more extensive \textsc{Wikispeedia} dataset against the WCM dense graph have also shown that the top metrics were measured on the WCM dense graph, despite its smaller size. This indicates that the model's success is influenced by the quality of the graph's structure and the features used.

\section*{Acknowledgements}

This research was carried out as part of the project ``Optimal Path Recommendation with Multi Criteria” (Project code: KMP6-0078997) under the framework of the Action ``Investment Plans of Innovation” of the Operational Program ``Central Macedonia 2014-2020” that is co-funded by the European Regional Development Fund and Greece.

\bibliographystyle{IEEEtran}
\bibliography{strings}

\end{document}